\documentclass[11pt]{article}
\usepackage{epsf}
\usepackage{slashed}
\usepackage{color}
\begin{document}
\begin{titlepage}
\title{Revisiting Weak Radiative Decays of Hyperons}
\author{
{Piotr \.Zenczykowski }\footnote{E-mail: piotr.zenczykowski@ifj.edu.pl}\\
{\em Division of Theoretical Physics}\\
{\em the Henryk Niewodnicza\'nski Institute of Nuclear Physics}\\
{\em Polish Academy of Sciences}\\
{\em Radzikowskiego 152,
31-342 Krak\'ow, Poland}\\
%E-mail: piotr.zenczykowski@.ifj.edu.pl
}
\maketitle
\begin{abstract}
Triggered by experimentally driven renewed interest in hyperon properties we address
the subject of weak radiative hyperon decays (WRHD). We start with the issue of Hara's theorem and briefly discuss the question of its possible evasion. Then, we give a short review of the story of  vector-meson-dominance (VMD) approach to  WRHD. We stress the shift from the Hara's-theorem-violating to Hara's-theorem-satisfying version of the VMD approach that did occur over time. Finally, spurred by a recent theoretical paper, we discuss the pole model description of WRHD, putting special attention to the issue of
the contributions from the intermediate $\Lambda(1405)$ state. We point out that
the measurement of the $\Lambda \to n \gamma$ decay asymmetry could resolve the encountered ambiguities and definitely answer the question of whether Hara's theorem is violated or not.
\end{abstract}

\vfill
\end{titlepage}

\section{Introduction}
\label{sec1}
The problem of weak radiative hyperon decays (WRHD) has been with us for some 60 years.
It may be regarded as a weak-interaction-involving counterpart of the 
issue of baryon magnetic moments. Just as understanding of the latter gave
us important information on the electromagnetic interaction of strongly interacting particles,
WRHD present us with unique and very simple probe on their joint electromagnetic and weak interaction. The two problems look very similar and simple as far as strong interactions are concerned:
in both processes  there are only one incoming and one outgoing hadron 
 (plus the photon), thus maximally reducing any complications possibly induced by strong interactions.
Yet, while the issue of baryon magnetic moments has been sufficiently well understood right from
the very beginning of the quark model, the problem of WRHD still stirrs considerable controversy.
In effect it is being approached by
subsequent generations of physicists again and again, with their conclusions
differing conceptually (and numerically) in substantial ways.\\ 

Recently collected BESIII data on the $J/\psi$
production and its decay into hyperons provide an incentive to readdress the issue
of WRHD. Spurred by these experimental developments, a new round of discussion on WRHD is
to be expected (see eg. \cite{HBLi,JMRChina,ChinaSU3}). In particular, paper \cite{JMRChina} adresses the issue in the
framework of the non-relativistic constituent quark model, used to evaluate the weak and electromagnetic transitions
relevant for their incorporation into the baryonic pole model. The authors of paper \cite{JMRChina} argue that the $1/2^-$
pole terms provide a natural mechanism for evading Hara's theorem \cite{Hara}, thus explaining the
large negative asymmetry observed in the $\Sigma^+ \to p \gamma$ decay. As it is the
issue of Hara's theorem that divides the (often very respectable) authors of various WRHD papers into two opposing camps (ie. accepting or rejecting the theorem), the present paper begins with
a presentation of Hara's theorem and the assumptions it is based on (Section \ref{HT}). Then, in Section \ref{hist}, the story of the vector-meson dominance (VMD) approach to WRHD is briefly reviewed, ending with a brief  
description explaining my shift (still within the general VMD framework) from the `against-Hara' to the `pro-Hara' camp. 
In the subsequent Section \ref{disc} the results of the old VMD and some quark model calculations are compared with those of ref. \cite{JMRChina}. In this Section a thorough discussion and our views on the
results of \cite{JMRChina} are also presented. In particular, it is pointed out that 
 the observed negative sign of the (experimentally sizable) $\Xi^0 \to \Lambda \gamma$ asymmetry 
(a symmetry-related counterpart of $\Sigma^+ \to p \gamma$)
does not constitute a telltale and unquestionable sign of a Hara's theorem satisfying approach.
Nonetheless, we stress that -- while the $\Xi^0 \to \Lambda \gamma$ asymmetry may turn out to be negative in some Hara's theorem violating calculations -- there exist important  experimental and theoretical arguments that strongly support the theorem. Finally, we point out that measurement of the $\Lambda \to n \gamma$ asymmetry should yield a definite answer on the issue of Hara's theorem.

\section{Hara's theorem}
\label{HT}
Hara's theorem \cite{Hara} is concerned with the properties of the parity-violating (p.v.) $\Sigma^+ \to p \gamma$
decay amplitude. Being originally formulated  in the language of local field theory at hadronic level, it assumes electromagnetic gauge invariance and CP-invariance. Under these two unshakeable assumptions it states that the said p.v. amplitude has to vanish in the limit of exact flavour SU(3). Given the fact that the mass of strange quark is in fact larger than that of the up and down quarks,
the theorem is clearly concerned with an unphysical limit (i.e. when the mass of $\Sigma^+$ becomes equal to that of proton). 
Yet, the case of baryon magnetic moments which are fairly well described within the SU(3) symmetric framework (with only some 20\% observable deviations from that limit) strongly suggests that
this unphysical SU(3)-symmetry limit should give predictions for WRHD that are not far from reality. Thus, assuming that the parity-conserving (p.c.)
$\Sigma^+ \to p \gamma$ amplitude is substantial (as suggested by the sizable $\Sigma^+ \to p \gamma$ branching ratio) one would expect the $\Sigma^+ \to p \gamma$
asymmetry to be small (say, of the order of 20\% of the maximal allowed asymmetry of $\pm 1$). Instead, experiment shows that
the said asymmetry is large (and negative: $\alpha_{exp}(\Sigma^+ \to p \gamma)=-0.76 \pm 0.08$ \cite{PDG}), implying that both p.c. and p.v. amplitudes are substantial.\\

This situation and various theoretical calculations (such as eg. \cite{KR,PZVMD1}) 
may be taken to suggest that
Hara's theorem does not hold in the proper approach to WRHD. 
Naturally, if Hara's theorem is not true, at least one of its assumptions must be incorrect.
The problem is that these assumptions (CP-invariance, electromagnetic gauge invariance, and hadron-level locality) are fundamental and virtually untouchable. Indeed,
the theorem follows directly from the consideration of the only parity-violating 
$\Sigma^+ p \gamma$ coupling
that is permitted  by CP-conservation and gauge invariance in hadron-level theoretical language, ie. from
\begin{equation}
\label{Hara}
\left[\bar{\psi}_pi\sigma_{\mu\nu}\gamma_5\psi_{\Sigma^+}-
\bar{\psi}_{\Sigma^+}i\sigma_{\mu\nu}\gamma_5\psi_{p}\right]q^{\mu}A^{\nu}.
\end{equation}
Now,
the weak Hamiltonian is symmetric under the $s\leftrightarrow d$ interchange. 
Moreover, under 
this interchange one has
$\Sigma^+(uus) \leftrightarrow p (uud)$.
In the case of exact SU(3), when the masses of $\Sigma^+$ and $p$ become identical, the expression  relevant for the description of the p.v. $\Sigma^+ \to p \gamma$
amplitude should therefore be completely symmetric under   $\Sigma^+(uus) \leftrightarrow p (uud)$.
Since expression (\ref{Hara}) is antisymmetric, and a symmetric part of an antisymmetric expression is zero, Hara's theorem immediately follows.
Further details of the proof of the theorem may be found eg. in \cite{LachZen}.\\

Since the size of  the experimental $\Sigma^+ \to p \gamma$ branching ratio is of the right
order for a CP-conserving process, evasion of Hara's theorem may follow only from
modifications to locality and/or electromagnetic gauge invariance. Both options seem to
require unorthodox views concerning the concept of spacetime point. 
For example, the origin of the violation of Hara's theorem in the quark model
of Kamal and Riazuddin (KR) \cite{KR} lies in the fact that in this calculation quarks are described by plane waves that are not really confined.
The hadron-level description used in \cite{KR} is that of a {\it multi}-local 
field theory, describable in position space by $\psi (x_1,x_2,x_3)$ ($x_k$ being $k$-th quark location), with 
{\it no restrictions} on interquark distances
$|x_m-x_n|$. Such a framework
is not reducible to the language of an effective hadron-level local field theory
used in the proof of Hara's theorem. In the KR scheme the quarks are free and do not feel the effects
of confinement. In other words, the KR calculations involve severe nonlocality that provides the reason for the violation of Hara's theorem.\\

The argument that an explanation of Hara's theorem violation could be attributed to the  point-like nature of hadrons being only approximate (but not severely nonlocal in the sense of \cite{KR}) should not be expected to work, as Hara's theorem deals with the limit of vanishing photon momentum.
In this very long wavelength limit the spatial internal structure of hadrons should cease to be discernible and the hadronic behaviour
should be satisfactorily described in the language of an {\it effective} local field theory at hadron level. Yet, various explicit quark-based calculations indicate violation of the theorem.

\section{A brief history of the VMD approach}
\label{hist}
A somewhat different scheme that suggests evasion of Hara's theorem is the calculation of \cite{PZVMD1}
which is based on the idea of vector meson dominance (VMD). According to VMD the coupling of photon to hadrons is proportional to an appropriate vector-meson-hadron coupling, with the factor of proportionality
being basically the ratio of electric charge and a strong hadron-level vector-meson-hadron coupling. The rationale for the application of VMD in \cite{PZVMD1} was simple: it followed from the general acceptation of the view that `VMD always works'. It appears then that
when the parity-violating couplings of vector mesons to baryons are assumed to be of the form
derived in simple quark model/$SU(6)_W$ studies on nuclear parity violation \cite{DDH}, the application of the idea of VMD
leads to the violation of Hara's theorem. The origin of this result stems
from the form of vector-meson-baryon-baryon parity-violating $B_i\to B_fV$ couplings, identified in \cite{DDH} with  
\begin{equation}
\label{gmg5}
V^{\mu}\bar{\psi}_f\gamma_{\mu}\gamma_5\psi_i.
\end{equation}
Using VMD, Eq. (\ref{gmg5}) suggests  the existence of a nonvanishing $A^{\mu}\bar{\psi}_f\gamma_{\mu}\gamma_5\psi_i$ photon-baryon term. If such a term does exist in addition to the standard $A^{\mu}\bar{\psi}_f\sigma_{\mu\nu}\gamma_5\psi_{i}$ coupling of Eq. (\ref{Hara}), the assumptions of Hara's theorem are
not satisfied and the theorem could be violated. The problem is that such a term violates electromagnetic gauge invariance at hadronic level
 (see eg. \cite{LachZen}). Thus,
if such an effective term exists in the real world, something very peculiar must be going on.\\

Instead of addressing directly the theoretical aspects of the puzzle raised by the suggested existence of the $A^{\mu}\bar{\psi}_f\gamma_{\mu}\gamma_5\psi_i$ coupling, one can look  for experimental clues that could help with its solution.  Indeed,
by analogy with the issue of baryon magnetic moments one can expect that experimental data 
on $\Sigma^+ \to p\gamma$ and its symmetry-related WRHD counterparts (ie. $\Sigma^0 \to n\gamma$,
$\Lambda \to n \gamma$, $\Xi^0\to\Lambda\gamma$, $\Xi^0 \to \Sigma^0\gamma$, and 
$\Xi^-\to \Sigma^-\gamma$) should not deviate far from the SU(3) limit. Thus, if the Hara's violating
and Hara's satisfying alternatives lead to two sets of well distinguishable predictions for other 
weak radiative parity-violating amplitudes, experiment could give us important hints
on the theoretical issue of what happens in the SU(3) limit.\\

Since the very small size of the experimental $\Xi^- \to \Sigma^- \gamma$ branching ratio tells us that the $s \to d \gamma$ single quark transition is negligible, the dominant part of weak transition should involve $W$-exchange between quarks.
 Furthermore, one should
be concerned mainly with its contribution to the p.v. WRHD amplitudes (as no substantial controversies surface in the p.c. amplitudes). There are two possible time orderings
in which the interquark $W$-exchange and the photon emission may occur. In the SU(3)-symmetric approach and 
for each one of the two orderings separately  the contributions to the 
amplitudes of the $\Sigma^+ \to p\gamma$, $\Sigma^0 \to n\gamma$,
$\Lambda \to n \gamma$, $\Xi^0\to\Lambda\gamma$, and $\Xi^0 \to \Sigma^0\gamma$ decays
are related by SU(3).
 There remains the question of how the amplitudes corresponding to these two orderings should be combined with one another, ie. should they be added or subtracted (for relatively real amplitudes).
This involves the issue of the symmetry properties of the whole amplitude under $i \leftrightarrow f$
interchange. One can convince oneself (see eg. \cite{LachZen}) that the 
$A^{\mu}\bar{\psi}_f\sigma_{\mu\nu}\gamma_5\psi_{i}$ 
($A^{\mu}\bar{\psi}_f\gamma_{\mu}\gamma_5\psi_i$) expressions
correspond to the subtraction (addition) of the amplitudes relevant for the two
time orderings (the $A$ and $B$ amplitudes of \cite{JMRChina}). 
Thus, if the two resulting possibilities for the total p.v.
amplitudes markedly differ, data on branching ratios and asymmetries may resolve
the issue of the violation of Hara's theorem. 
The set of relevant $B_i \to B_fV$ baryon-baryon-vector-meson SU(3)-symmetric $A$- and $B$- type amplitudes is given in Table \ref{tableWRHDpz1}
(it is taken from \cite{PZVMD3} with amplitude signs adjusted to fit those of the p.c. amplitudes from the recent work of \cite{JMRChina}.)
Hara's theorem
appears in the subtraction prescription as a result of the cancellation of the identical coefficients ($-1/3\sqrt{2}$) at the $A$- and $B$- type  amplitudes
contributing to $\Sigma^+ \to p \gamma$. 
In other words, the relative scale of the $A$ and $B$ amplitudes is fixed in the SU(3) symmetry limit. Detailed calculations (eg. in \cite{Orsay})
show that the relevant amplitude is $+(A-B)$ (ie. not $-(A-B)$).
Thanks to the essential differences between the subtraction and addition alternatives 
the signs of some WRHD asymmetries appear to be of particular significance (see Table \ref{tableWRHDpz1}).
In fact, it was argued in \cite{LachZen} that the crucial experimental
number is the sign (and size) of the $\Xi^0 \to \Lambda \gamma$ decay 
asymmetry.\footnote{As can be seen from Table \ref{tableWRHDpz1} there is another Hara's-theorem-sensitive asymmetry, namely that of $\Lambda \to n \gamma$. Yet, as this asymmetry is very hard to be determined experimentally, no stress was put in the past on the importance of its measurement.} 
Actually, this conclusion is
valid 
provided a `sufficiently symmetric description' 
of the p.v. amplitudes is adopted.
The meaning of the term `sufficiently symmetric' will
become clear at the end of this Section.
 \\

\begin{table}[h]
\caption{Parity-conserving and parity-violating $B_i \to B_f U_0$ baryon-baryon-vector-meson amplitudes (in units of $10^{-7}$; $b_R=5.3$) \cite{PZVMD3}. $U_0$ denotes $U$-spin-$0$ vector meson. The p.v. amplitudes, adapted from \cite{PZVMD3}, respect SU(3) symmetry. Amplitudes $A$ and $B$ correspond to the $A$ and $B$ pieces of the p.v. WRHD amplitudes as discussed in \cite{JMRChina}. Columns 4 and 5 contain the coefficients at $b_R$ in the $A$ and $B$ amplitudes respectively. In the last two columns the signs of asymmetries for the
Hara's satisfying (violating) combinations of the $A$ and $B$ amplitudes are given.  
The signs of all amplitudes are adjusted to fit the signs of p.c. amplitudes in Table II of
\cite{JMRChina}.
}
\begin{center}
\begin{tabular}{ccccccc}\hline
Decay & tot p.c. & tot p.v. ($A-B$) &  p.v. $A$ & p.v. $B$ & \multicolumn{2}{c}{ asym.  signs}\rule{0mm}{6mm}\\
&&&&&$A-B$&$A+B$\rule{0mm}{6mm}\\
&&&&&(Hara)&(no~Hara)\rule{0mm}{6mm}\\
\hline
$\Sigma^+\to p U^0$       & $+18.8$  &$(-\frac{1}{3\sqrt{2}}+\frac{1}{3\sqrt{2}})b_R$ &  $-\frac{1}{3\sqrt{2}}$  & $-\frac{1}{3\sqrt{2}}$ &  $0$ & $-$\rule{0mm}{6mm}\\
%\hline
$\Sigma^0\to nU^0$        & $+42.1$  &$(+\frac{1}{6}-\frac{1}{6})b_R$ & $+\frac{1}{6}$ & $+\frac{1}{6}$
&$0$&$+$\rule{0mm}{6mm}\\
%\hline
$\Lambda \to n U^0$       & $+15.7$  & 
$(+\frac{1}{6\sqrt{3}}-\frac{1}{2\sqrt{3}})b_R$ 
& $+\frac{1}{6\sqrt{3}}$
&$+\frac{1}{2\sqrt{3}}$
&$-$&$+$\rule{0mm}{6mm}\\
%\hline
$\Xi^0\to \Lambda U^0$    & $-13.9$  & $(0+\frac{1}{3\sqrt{3}})b_R 
$ &$0$
& $-\frac{1}{3\sqrt{3}}$ & $-$ &$+$\rule{0mm}{6mm}\\
%\hline
$\Xi^0\to \Sigma^0 U^0$   & $-62.1$  &$(+\frac{1}{3}-0)b_R$
&$+\frac{1}{3}$
&$0$& $-$ &$-$\rule{0mm}{6mm}\\
\hline
\end{tabular}
\end{center}
\label{tableWRHDpz1}
\end{table}

It was stressed in \cite{LachZen} that the available description of the parity-conserving $\Xi^0 \to \Lambda \gamma$ is very reliable and that, for sufficiently symmetric descriptions of the p.v. amplitudes (such as those of eg. \cite{KR,PZVMD1,Orsay}, see  later for a more detailed clarification), the size and sign of
 the $\Xi^0 \to \Lambda \gamma$ asymmetry are predicted  to be large ($0.7 -0.9$ in absolute value)
and negative (positive) for the Hara's theorem satisfying (violating) case.
The difference in sign can be readily traced to the difference in sign between
the $A\pm B$ options, which follows from the fact that in the appropriate symmetry limit the relevant total $A$-type $\Xi^0 \to \Lambda \gamma$ amplitude
is zero. 
Thus, when Borasoy and Holstein claimed in their paper \cite{BH} that in the ChPT (Hara's theorem satisfying) approach
one gets $\alpha(\Xi^0 \to \Lambda \gamma)=+0.46$, one could suspect that their paper contains an error. Indeed, it was soon shown \cite{PZcomBH} that the approach
of \cite{BH} omits the contribution from the intermediate $1/2^-$  SU(3)-singlet state
(i.e. from $\Lambda(1405)$). When this contribution is added as in \cite{Orsay} (ie. in a sufficiently symmetric way, see later) one recovers the negative sign and significant size of the $\Xi^0 \to \Lambda \gamma$ asymmetry \cite{PZcomBH}.\\

The existence of the problem with the treatment of the contribution from the intermediate $\Lambda(1405)$ may be conjectured from the consideration of the decomposition of the $A$- and $B$- type p.v. amplitudes 
of Table \ref{tableWRHDpz1}
into the individual
contributions from the intermediate $J^P=1/2^-$ states of $^28$, $^48$, and $^21$  from the ($70,1^-$) multiplet of $SU(6)\times O(3)$.
This decomposition  is given
in Table \ref{tableWRHDpzBH} (adapted from \cite{PZcomBH} to fit the p.c. phases in Table
\ref{tableWRHDpz1} and in \cite{JMRChina}). 
The coefficients at the $A$ and $B$-type amplitudes of Table \ref{tableWRHDpz1} are proportional 
to the relevant entries (marked {\it ``all''}) in Table \ref{tableWRHDpzBH}
(by construction, with positive proportionality sign, ie. $b_R \propto 2+K$). 
\footnote{The $K$ parameter was estimated in \cite{PZcomBH} to be around $1$ (for ref.\cite{Orsay} one finds $K\approx \omega/m \approx 1.25$, with $\omega$ being h.o. excitation frequency in the constituent quark model, and $m$ being SU(3)-symmetric constituent quark mass).} 
Thus, simple quark model /VMD expressions are obtained through the summation over
the individual contributions from the $1/2^-$ intermediate states. In other words, the relative signs
of the individual pole model contributions may be easily cross-checked from the condition that (in the appropriate symmetry case) they should add up to the simple quark model/VMD results. When summed up over all pole terms, the overall pattern of pole model contributions (sometimes regarded as
long-distance contributions) must reproduce the pattern of the
quark model/VMD calculations (separately for the $A$- and $B$-type amplitudes). Any distinction between the quark-level and baryon-level terms should be irrelevant as far as effective SU(3) properties of the amplitudes are concerned. 
From Table \ref{tableWRHDpzBH} it is immediately seen that the contribution from $^21$ is essential for the appearance of the vanishing total
$A$-type $\Xi^0 \to \Lambda \gamma$ amplitude.
As shown in \cite{PZcomBH} in more detail, when this contribution from $\Lambda(1405)$ is taken into account 
in a sufficiently symmetric way
one tends to obtain a sizable negative $\Xi^0 \to \Lambda \gamma$ asymmetry, thus
reproducing  the predictions of \cite{Orsay} quite well.\\

\begin{table}[hp]
\caption{Decomposition of p.v. SU(3)-symmetric $A$ and $B$ amplitudes into contributions from the ($70,1/2^-$) intermediate states (adapted from \cite{PZcomBH} to fit the phases in Table
\ref{tableWRHDpz1}). For the value of $K$ see text.}
\begin{center}
\begin{tabular}{cccccc}\hline
Decay   &   int. state & p.v. $A$ & p.v. $B$ & $A+B$ & $A-B$ \rule{0mm}{6mm}\\
\hline
$\Sigma^+\to p \gamma$          
&  $^28$  & $-\frac{1}{3\sqrt{2}}(2+K)$ & $-\frac{1}{3\sqrt{2}}(2+K)$ &&\rule{0mm}{6mm}\\
&$^48$&$0$&$0$&&\rule{0mm}{6mm}\\
&$all$&$-\frac{1}{3\sqrt{2}}(2+K)$ & $-\frac{1}{3\sqrt{2}}(2+K)$ &$-\frac{2}{3\sqrt{2}}(2+K)$& $0$\rule{0mm}{6mm}\\
\hline
$\Lambda \to n \gamma$       
&  $^28$&$\frac{1}{6\sqrt{3}}(2+\frac{K}{3})$&$\frac{1}{3\sqrt{3}}(2+\frac{K}{3})$&&\rule{0mm}{6mm}\\
&$^48$&$\frac{1}{9\sqrt{3}}K$& $\frac{2}{9\sqrt{3}}K$ &&\rule{0mm}{6mm}\\
&$^21$&$0$&$\frac{1}{6\sqrt{3}}(2+K)$&&\rule{0mm}{6mm}\\
&$all$&$\frac{1}{6\sqrt{3}}(2+K)$&$\frac{1}{2\sqrt{3}}(2+K)$&$\frac{2}{3\sqrt{3}}(2+K)$&$-\frac{1}{3\sqrt{3}}(2+K)$\rule{0mm}{6mm}\\
\hline
$\Xi^0\to \Lambda \gamma$      
&$^28$
& $-\frac{1}{6\sqrt{3}}(2+\frac{K}{3})$ &  $-\frac{1}{3\sqrt{3}}(2+\frac{K}{3})$ & &\rule{0mm}{6mm}\\
&$^48$&$-\frac{1}{9\sqrt{3}}K$&$-\frac{2}{9\sqrt{3}}K$&&\rule{0mm}{6mm}\\
&$^21$&$\frac{1}{6\sqrt{3}}(2+K)$&$0$&&\rule{0mm}{6mm}\\
&$all$&$0$&$-\frac{1}{3\sqrt{3}}(2+K)$&$-\frac{1}{3\sqrt{3}}(2+K)$
&$\frac{1}{3\sqrt{3}}(2+K)$\rule{0mm}{6mm}\\
\hline
$\Xi^0\to \Sigma^0 \gamma$      
&$^28$&$\frac{1}{6}(2+\frac{K}{3})$&$0$&&\rule{0mm}{6mm}\\
&$^48$&$\frac{1}{9}K$&$0$&&\rule{0mm}{6mm}\\
&$^21$&$\frac{1}{6}(2+K)$&$0$&&\rule{0mm}{6mm}\\
&$all$&$\frac{1}{3}(2+K)$&$0$&$\frac{1}{3}(2+K)$&$\frac{1}{3}(2+K)$\rule{0mm}{6mm}\\
\hline
\end{tabular}
\end{center}
\label{tableWRHDpzBH}
\end{table}

At the time ref. \cite{LachZen} was written and for a couple of years afterwards, experiment indicated substantial positive
asymmetry of the $\Xi^0 \to \Lambda \gamma$ decay, supporting the belief that Hara's theorem is violated (my last paper adopting this view being \cite{PZ2002}). 
It was only around the time of the publication of \cite{PZ2002} that the NA48 experiment \cite{NA48} 
measured the   
$\Xi^0 \to \Lambda \gamma$ asymmetry to be $-0.78 \pm 0.19$ ie. large and negative, thus providing a strong experimental argument for Hara's theorem being satisfied. This triggered 
my shift from the against-Hara to the pro-Hara camp.
In order
to agree with the new data the subsequent VMD papers \cite{PZVMD3,PZVMD2}
accepted that the so-far used (quark-model-based) description of the vector-meson-baryon-baryon p.v. couplings (as given in
\cite{DDH})
has to be substantially modified. The relevant modification does not only explain the observed signs and absolute magnitudes of the WRHD but - at the same time - it resolves another
old problem in weak hyperon decays: the discrepancy between the $f$ and $d$ SU(3) coupling
 constants as observed in the $S$- and $P$- waves of nonleptonic hyperon decays (NLHD)
\cite{DGH}.
Specifically, \cite{PZVMD3,PZVMD2} explain why, contrary to the soft-meson theorems 
predicting $f_P/f_S=d_P/d_S=1$, one has
$f_P/f_S \approx 1.5$ and $d_P/d_S \approx 2.2$. 
Given the simultaneous resolution of the problems apppearing in NLHD and WRHD one has
to regard the Hara's theorem satisfying (SU(3) breaking) approach of \cite{PZVMD3} as the most likely resolution of
the relevant problems. \\

As far as details are concerned, paper \cite{PZVMD3} employs
a `sufficiently symmetric' $1/2^-$ pole-model description of the parity-violating amplitudes
(as used in \cite{Orsay}).
In this description (the following clarifies the idea of `sufficiently symmetric') one breaks SU(3) symmetry 
between the $A$ and $B$ amplitudes 
%(as given in Table \ref{tableWRHDpz1}) 
in a simplified way. First,  SU(3)-breaking is considered in the pole model denominators only. Second,
{\it all} $A$- amplitudes are made relatively larger by a {\it single} pole-model-induced denominator factor $\omega/(\omega -\Delta m_s)$ (and, likewise, all $B$-amplitudes are made relatively smaller by an analogous  factor $\omega/(\omega+\Delta m_s)$), with $\omega \approx 570~MeV$ being the h.o. excitation frequency and $\Delta m_s \approx 190~MeV$ being the strange-nonstrange quark mass difference. 
In other words, in such a `sufficiently symmetric' description the relative sizes of all $A$-type amplitudes (and, separately, the relative sizes of all $B$-type amplitudes) stay unchanged among themselves. Accordingly, the proportions of the contributions from the
$^28$, $^48$, and $^21$ multiplets stay unchanged within the whole $A$ (or $B$) amplitude group (ie. when compared
with those given in Table \ref{tableWRHDpzBH}).
Thus, for example,
for the  $\Xi^0 \to \Lambda \gamma$ decay
the $A$-type contributions from the $^28$, $^48$, and $^21$ states still add up to $0$
(as in Table \ref{tableWRHDpzBH}), and it is the $B$-type amplitude alone that determines
the relevant $\Xi^0 \to \Lambda \gamma$ asymmetry.\\

We conclude this section on the history of the application of VMD to
the description of WRHD by stressing that the final (Hara's-theorem satisfying) VMD papers 
\cite{PZVMD3,PZVMD2}
markedly differ from the earlier ones in which Hara's theorem is violated (eg. \cite{PZVMD1}):
the two groups of VMD papers use two completely different forms of the p.v. $B_i \to B_fV$ amplitudes (ie. $A+B$ in the first group and $A-B$ in the second group). 
\\

\section{Contributions of intermediate $\Lambda(1405)$}
\label{disc}
 {\small
\begin{table}[thp]
\caption{Weak radiative hyperon amplitudes as evaluated in \cite{JMRChina}. P.v.  amplitudes arising from the intermediate excited $\Lambda$ and $\Sigma$ states (separately from the $^28$ and the $^48$ multiplets) have been added.}
\begin{center}
\begin{tabular}{cccccc}
%\hline
 Decay     &   tot.~p.c.   &    tot.~p.v.    & int. ~st.&      p.v. $A$     &  p.v. $B$     \rule{0mm}{6mm}\\
\hline
$\Sigma^+\to p\gamma $    &  $5.10$    &  $-15.40-2.66i$  & $^28$& $-9.65-2.39i$ &  $ -5.75-0.27i$  \rule{0mm}{6mm}\\
    &      &       &     $^48$  & $0$ &  $0$    \rule{0mm}{6mm}\\
    &      &       &     $all$  & $-9.65-2.39i$ &  $ -5.75-0.27i$    \rule{0mm}{6mm}\\
 \hline
$\Sigma^0\to n \gamma$    &  $6.69$    &   $1.33+1.37i$    & $^28$ &   $7.69+1.92i$    &  
$-0.13-0.04i$    \rule{0mm}{6mm}\\
&&&$^48$&   $-0.68-0.15i$   &$0.02-0.05i$\rule{0mm}{6mm}\\
&&&$^21$&&  $-5.58-0.36i$\rule{0mm}{6mm}\\
&&&$all$&$7.01+1.77i$&  $-5.69-0.45i$\rule{0mm}{6mm}\\
\hline
$\Lambda\to n \gamma$    &  $5.82$    &   $-14.91-1.48i$    & $^28$ &   $-4.72-0.99i$    &  $-6.45-0.28i $    \rule{0mm}{6mm}\\
   &      &       &     $^48$  & $0.22+0.03i$ &  $0.31-0.02i $    \rule{0mm}{6mm}\\
   &      &       &     $^21$  &  & $-4.27-0.28i$      \rule{0mm}{6mm}\\
  &      &       &     $all$  & $-4.50-0.96i$ & $-10.41-0.58i$      \rule{0mm}{6mm}\\
\hline
$\Xi^0\to \Lambda \gamma$    &  $-7.81$    &   $-4.38-3.88i$    &  $^28$&   $4.76+0.26i$    & 
$6.33+0.22i$ \rule{0mm}{6mm}\\
&&&$^48$&$-0.23-0.08i$&$-0.46-0.01i$\rule{0mm}{6mm}\\
&&&$^21$&$-14.7-4.26i$&\rule{0mm}{6mm}\\
&&&$all$&$-10.17-4.08i$&$5.87+0.21i$\rule{0mm}{6mm}\\
\hline
$\Xi^0 \to \Sigma^0 \gamma$    &  $-8.15$    &   $-45.65-10.67i$    &  $^28$  & $-10.75-0.59i$ & $0$   \rule{0mm}{6mm}\\
&&&$^48$&$0.21+0.07i$&$0$\rule{0mm}{6mm}\\
&&&$^21$&$-35.11-10.15i$&\rule{0mm}{6mm}\\
&&&$all$&$-45.65-10.67i$&\rule{0mm}{6mm}\\
\hline
\end{tabular}
\end{center}
\label{tableWRHD}
\end{table}}
In a recent paper \cite{JMRChina} the issue of the pole model description of WRHD 
(considered as the dominant mechanism of these decays) was addressed anew. In the approach of \cite{JMRChina} the weak and electromagnetic transitions
involved in the description were evaluated within the SU(3)-symmetric constituent quark model. SU(3) breaking entered through the pole model denominators in which experimentally
observed $1/2^+$ and $1/2^-$ masses were used.
For  the purposes of our discussion, in Table \ref{tableWRHD} we list
the results of the calculations of the p.c. and p.v. amplitudes given in Table II of \cite{JMRChina}. 
When compared with the original Table II, our Table \ref{tableWRHD}
is simplified: we summed up the amplitudes arising from the intermediate $\Lambda$ and $\Sigma$
 for the $^28$ and (separately) for the $^48$ multiplets. In this way direct comparison
with our Table \ref{tableWRHDpzBH} becomes possible.\\

Joint inspection of Tables \ref{tableWRHDpzBH}, \ref{tableWRHD} reveals various similarities and differences, and permits drawing important conclusions.
Actually, a straightforward comparison of the two tables is not possible because
of different treatments of SU(3), which is exact in Table \ref{tableWRHDpzBH} but broken in
Table \ref{tableWRHD}.
Still, to the extent that SU(3) is not broken too much, the two tables should exhibit
important similarities.
Indeed, consider first the $\Sigma^+\to p\gamma$ p.v. amplitudes $A$ and $B$.
In both tables they are negative (our p.c. phases have been adjusted to fit those of \cite{JMRChina}). Furthermore, the $A$ amplitude in Table \ref{tableWRHD}
is larger in absolute magnitude than amplitude $B$, which agrees with the discussion
of the `sufficiently symmetric' SU(3)-breaking extension of Table \ref{tableWRHDpzBH}
(see the preceeding section),
according to which the scale of the $A$ ($B$) amplitudes becomes relatively larger (smaller)
than that given in Table \ref{tableWRHDpzBH}.\\

Upon closer inspection a similar enhancement (reduction) of the absolute magnitude of the $A$ ($B$) amplitudes can be seen for the remaining $\Lambda \to n \gamma$, $\Xi^0 \to \Lambda\gamma$, and $\Xi^0\to \Sigma^0\gamma $ decays. Consider for example the $^21$ contributions to the $\Lambda\to n \gamma$ and $\Xi^0\to \Lambda\gamma$ decays. In the SU(3)-symmetric case they should be equal in magnitudes (see Table \ref{tableWRHDpzBH}), but in the SU(3)-breaking case due to the energy denominator effects the $^21$ contribution to the $A$($\Xi^0\to \Lambda\gamma$) amplitude should be larger (in absolute magnitude) than the $^21$ contribution to the $B$($\Lambda\to n \gamma$) amplitude. This is well seen in Table \ref{tableWRHD} where the physical mass value of the $^21$ state (ie. $\Lambda(1405)$) is used. In fact, as calculated in \cite{JMRChina}, due to the low mass of $\Lambda(1405)$ the size of such effects may sometimes be very substantial. For example, for the p.v. $\Xi^0\to\Lambda\gamma$ amplitude in \cite{JMRChina} the $^21$ $A$-type contribution becomes larger than the sum of {\it all} the remaining ($A$ and $B$) terms. 
\footnote{\label{fn}
A simple estimate of this effect may be obtained by considering the ratio of the pole model energy denominators relevant for
$A(\Xi^0\to\Lambda\gamma)$ amplitudes with intermediate $^21$ and $^28$ states:
$(M_{\Lambda(1670)}-M_{\Xi(1310)})/(M_{\Lambda(1405)}-M_{\Xi(1310)}) \approx 3.5 $.
This could be compared with the ratio of $^21$ and $^28$ $A$-type 
$\Xi^0\to \Lambda\gamma$ p.v. amplitudes of Table \ref{tableWRHD} which is (roughly): $|-14.7/4.76| \approx 3 $.} \\

Still, there is an important difference between the $\Lambda \to n \gamma$, $\Xi^0 \to \Lambda\gamma$, and $\Xi^0\to \Sigma^0\gamma $ parity violating amplitudes of Table \ref{tableWRHDpzBH} and those of Table \ref{tableWRHD}: the dominant
($^28$ and $^21$) contributions in Table \ref{tableWRHD} differ in sign from those in Table \ref{tableWRHDpzBH}. As stressed in the previous section the relative signs 
of the $A$ and $B$ amplitudes are universal in the SU(3) limit and provide a useful cross-check on the calculations. Thus,
this overall sign difference constitutes a problem for the
$\Lambda \to n \gamma$, $\Xi^0 \to \Lambda\gamma$, and $\Xi^0\to \Sigma^0\gamma $ lines of 
Table II in \cite{JMRChina} (but not for the $\Sigma^+ \to p \gamma$ line).
  It would be therefore
worthwhile to see what happens with the entries of Table II of \cite{JMRChina} in the SU(3) limit, compare the individual $A$ and $B$ amplitudes with earlier calculations (eg. \cite{Orsay},\cite{PZcomBH}) and trace the origin of sign discrepancy. This is important
as the signs of $\Lambda\to n \gamma$ and $\Xi^0\to\Sigma^0\gamma$ asymmetries (as calculated
in \cite{JMRChina}) are opposite to the experimental ones, thus hinting quite clearly that there is a problem with the signs of the three ($\Lambda \to n \gamma$, $\Xi^0 \to \Lambda\gamma$, and $\Xi^0\to \Sigma^0\gamma $) p.v. amplitudes of \cite{JMRChina}. In fact, we will argue below that the signs of the $\Lambda\to n \gamma$ and $\Xi^0\to\Sigma^0\gamma$
asymmetries are much less model-dependent than the sign of the $\Xi^0 \to \Lambda\gamma$ asymmetry, thus further confirming our disbelief in the overall sign of the three relevant sets
of p.v. amplitudes.\\

In order to correct for the sign inconsistency observed between some of the corresponding p.v. amplitudes of Tables \ref{tableWRHD} and \ref{tableWRHDpzBH}, and for the sake of the subsequent discussion, we now multiply all relevant ($\Lambda \to n \gamma$, $\Xi^0 \to \Lambda\gamma$, and $\Xi^0\to \Sigma^0\gamma $) p.v. amplitudes of Table \ref{tableWRHD} by $-1$ as suggested by Table \ref{tableWRHDpzBH}.\footnote{ Naturally, I prefer to believe in my own calculations, especially as they agree with those of \cite{Orsay} and various other papers.} This modification changes the signs of all involved asymmetries (calculated in \cite{JMRChina} to be, respectively: $-0.67$, $+0.72$, and $+0.33$). Thus, it leads to
positive $\Lambda \to n \gamma $ asymmetry (ie. $+0.67$) and to negative asymmetries for
$\Xi^0 \to \Lambda\gamma$ and $\Xi^0\to \Sigma^0\gamma $ (respectively: $-0.72$ and $-0.33$). With the exception of 
$\Xi^0 \to \Lambda\gamma$ case this modification agrees with what was expected in the Hara's theorem violating case \cite{LachZen} (see asymmetry signs for the Hara's theorem violating ($A+B$) case in Table \ref{tableWRHDpz1}). It also agrees with experimental data that are available for $\Xi^0 \to \Lambda \gamma $ ($-0.70\pm 0.07$) and $\Xi^0 \to \Sigma^0 \gamma$ ($-0.69 \pm 0.06$) \cite{PDG}. The above procedure of sign reversal applied to the relevant amplitudes of Table \ref{tableWRHD} should not be considered as an `ad hoc' correction of
the observed problem of signs.
Instead, it should be viewed as an estimate of what would have happened in the Hara's theorem {\it violating} (VMD) version \cite{PZVMD1} of the pole model of \cite{PZVMD3} if the condition
of the sufficiently symmetric treatment of intermediate states were dropped and the physical masses of these states were used (see footnote \ref{fn} and the subsequent paragraph). \\

The issue of the negative sign obtained in this way for the $\Xi^0 \to \Lambda\gamma$ asymmetry
appears very interesting. Indeed, in Hara's theorem-violating and `sufficiently symmetric' case (compare Table  \ref{tableWRHDpzBH}) the $A$-type contributions
from the $^28$, $^48$, and $^21$ intermediate states add up to zero, in effect leading to a positive
$\Xi^0 \to \Lambda\gamma$ asymmetry. In the now discussed (sign-altered) modification of \cite{JMRChina}
the substantial negative $\Xi^0 \to \Lambda\gamma$ asymmetry appears because (as already discussed) the contribution from the intermediate $^21$ (ie. from $\Lambda(1405)$)
exceeds the sum of all other contributions  and reverses the expected sign of the total 
$A+B$ amplitude. Thus, we have a mechanism here (a highly dominant contribution from $\Lambda(1405)$, due to the smallness of its mass) which leads to a negative $\Xi^0 \to \Lambda\gamma$ asymmetry, as observed in the data.
In other words, paper \cite{JMRChina} (or, more precisely, its sign-modified version discussed here) brings attention to the fact that one can get a negative $\Xi^0 \to \Lambda\gamma$ asymmetry in the Hara's theorem violating case provided the symmetry of the spectrum of the intermediate $1/2^-$ states is severely broken. Therefore,
the negative $\Xi^0 \to \Lambda\gamma$ asymmetry, earlier argued \cite{LachZen} to be crucial, does not  constitute
an unquestionable sign of Hara's theorem being satisfied. Yet,
this concerns the contributions to the $A(\Xi^0 \to \Lambda\gamma)$ amplitude only. 
As can be seen from an inspection of the relevant entries in Tables \ref{tableWRHDpzBH} and \ref{tableWRHD}, a small value of the $\Lambda(1405)$ mass cannot change the predicted signs of the $\Lambda \to n \gamma $ and/or
$\Xi^0 \to \Sigma^0 \gamma$ asymmetries:
the relevant contributions from the intermediate $^21$ and $^28$ states add up constructively. 
In particular, in the case of Hara's theorem violation the $\Lambda \to n \gamma $ asymmetry is still predicted to be positive. Thus, measurement of this asymmetry would be extremely illuminating.
\\

\section{Conclusions}
Are we now back in the situation when Hara's theorem violation becomes a possible option?
In principle yes, with the problem certainly requiring a further study. However, it is hard to
believe today in the evasion of Hara's theorem. The reason is that the model discussed
in \cite{PZVMD3,PZVMD2} supplies a successful {\it unified} picture of both nonleptonic and radiative
weak hyperon decays, linking them together and explaining simultaneously both 1) the $S:P$ puzzle
in NLHD (ie. the sizes of the ratios of relevant SU(3)-invariant couplings $f_P/f_S$ and $d_P/d_S$ ) and 2) the set of experimental WRHD asymmetries and branching ratios in the orthodox (Hara's
theorem satisfying) approach. To the contrary, such a unified and parsimonious picture of NLHD and WRHD is so far absent
in the Hara's theorem violating case.\\
 
If one insists nonetheless on disregarding information coming from NLHD and restricts the considered experimental input to the WRHD data only, the resolution of the issue of Hara's theorem could come from the measurement of the $\Lambda \to n \gamma$
asymmetry. This asymmetry is so far unknown, but should be definitely negative (positive) in the Hara's theorem satisfying (violating) case as discussed at the end of the previous Section. Unfortunately, this decay presents severe problems on the experimental side.\\

\vfill

\vfill

\end{document}